\documentclass[twocolumn]{aastex62}

\graphicspath{{./}{figs/}}

\received{xxx}
\revised{xxx}
\accepted{xxx}
\submitjournal{ApJ}

\shorttitle{r-II Stellar Halo Stars from Destroyed Ultra-Faint Dwarfs}
\shortauthors{Brauer et al.}

\begin{document}

\title{The Origin of $r$-process Enhanced Metal-Poor Halo Stars In Now-Destroyed Ultra-Faint Dwarf Galaxies}

\correspondingauthor{Kaley Brauer}
\email{kbrauer@mit.edu}

\author{Kaley Brauer}
\affiliation{Department of Physics and Kavli Institute for Astrophysics and Space Research, Massachusetts Institute of Technology, Cambridge, MA 02139, USA}
\affiliation{Joint Institute for Nuclear Astrophysics -- Center for Evolution of the Elements, USA}

\author{Alexander P. Ji}
\altaffiliation{Hubble Fellow}
\affiliation{Observatories of the Carnegie Institution for Science, 813 Santa Barbara St., Pasadena, CA 91101, USA}
\affiliation{Joint Institute for Nuclear Astrophysics -- Center for Evolution of the Elements, USA}

\author{Anna Frebel}
\affiliation{Department of Physics and Kavli Institute for Astrophysics and Space Research, Massachusetts Institute of Technology, Cambridge, MA 02139, USA}
\affiliation{Joint Institute for Nuclear Astrophysics -- Center for Evolution of the Elements, USA}


\author{Gregory A. Dooley}
\affiliation{Google, 111 8th Ave, New York, NY 10011, USA}

\author{Facundo A. G\'omez}
\affiliation{Instituto de Investigaci\'on Multidisciplinar en Ciencia y Tecnolog\'ia, Universidad de La Serena, Ra\'ul Bitr\'an 1305, La Serena, Chile}
\affiliation{Departamento de F\'isica y Astronom\'ia, Universidad de La Serena, Av. Juan Cisternas 1200 N, La Serena, Chile}

\author{Brian W. O'Shea}
\affiliation{Department of Computational Mathematics, Science and Engineering, Michigan State University, MI, 48823, USA}
\affiliation{Department of Physics and Astronomy, Michigan State University, MI, 48823, USA}
\affiliation{Joint Institute for Nuclear Astrophysics -- Center for Evolution of the Elements, USA}



\begin{abstract}

The highly $r$-process enhanced (r-II) metal-poor halo stars we observe today could play a key role in understanding early ultra-faint dwarf galaxies, the smallest building blocks of the Milky Way. If a significant fraction of metal-poor r-II halo stars originated in the ultra-faint dwarf galaxies that merged to help form the Milky Way, observations of r-II stars could help us study these now-destroyed systems and probe the formation history of our Galaxy. To conduct our initial investigation into this possible connection, we use high-resolution cosmological simulations of Milky-Way-mass galaxies from the \textit{Caterpillar} suite in combination with a simple, empirically motivated treatment of $r$-process enrichment. We determine the fraction of metal-poor halo stars that could have formed from highly $r$-process enhanced gas in now-destroyed low-mass ultra-faint dwarf galaxies, the simulated r-II fraction, and compare it to the ``as observed" r-II fraction. We find that the simulated fraction, $f_{r-II,sim}\sim 1-2$\%, can account for around half of the ``as observed" fraction, $f_{r-II,obs}\sim 2-4$\%. The ``as observed" fraction likely overrepresents the fraction of r-II stars due to incomplete sampling, though, meaning $f_{r-II,sim}$ likely accounts for more than half of the true $f_{r-II,obs}$. Further considering some parameter variations and scatter between individual simulations, the simulated fraction can account for around $20-80$\% of the ``as observed" fraction. 

\end{abstract}



\section{Introduction} \label{sec:intro}

In the favored cosmological paradigm, galaxies grow hierarchically over time \citep{White78,Davis85}. Dark matter halos (and the galaxies inside them) merge together to form larger and larger galaxies, resulting in a final galaxy comprising both stars that formed \emph{in situ} and stars that formed in the now-destroyed progenitor galaxies. The \emph{in situ} stars are found primarily in the disk and bulge of the galaxy, where star formation is ongoing. The accreted stars are found primarily in the extended outskirts of the galaxy: the stellar halo \citep{Bell08}. The stars in a galaxy's stellar halo thus preserve information about the now-destroyed building blocks of that galaxy \citep{Bullock05}. 

The stellar halo can include a significant number of \textit{in situ} stars as well, though \citep{Monachesi18}. Furthermore, even among the stars that are believed to have been accreted, the properties of the galaxies in which they formed are largely a mystery. To decode the information stored in stellar halo stars, we must identify the stars that were accreted and determine the types of galaxies from which they accreted. One way to do this is by looking for stars with kinematic signatures of accretion \citep{Johnston96,Johnston98,Helmi99,Venn04}. Many galaxy mergers occurred early in the history of the galaxy, however, and by the time we observe the stellar halo, many of these kinematic signatures can be difficult to observe.

Selecting stars with specific chemical signatures provides another way forward. The Milky Way's accreted stellar halo is composed of long-ago destroyed galaxies covering a wide range of stellar masses. Those disrupted galaxies formed their stars at different rates, imprinting different chemical signatures on their most metal-poor stars \citep[e.g.,][]{Kirby11,DuaneThesis,Ishimaru15}. In particular, early $r$-process (rapid neutron-capture process) nucleosynthesis events in small dwarf galaxies would imprint a clean $r$-process signature on the subsequently formed stars in those galaxies.

The $r$-process is responsible for producing \textbf{around half of the abundances of} the heaviest elements in the periodic table \citep{Burbidge57,Cameron57}. For more information, see the recent review papers by \citet{Frebel18}, \citet{Thielemann17}, and \citet{2017PrPNP..94....1A}. Recently, it has become apparent that the majority of $r$-process material in the universe is likely synthesized in neutron star mergers (NSMs) \citep{Hotokezaka15,Ji16b,LIGOGW170817a,LIGOGW170817b}. Because neutron star mergers appear to have a long coalescence timescale (${\gtrsim}100$ Myr) and the metallicity of a stellar system increases with each new stellar generation, NSM events should only result in metal-poor stars with an $r$-process signature if those stars formed in dwarf galaxies with low star formation efficiencies (i.e., galaxies that form new generations of stars slowly relative to larger galaxies like the Milky Way or LMC) \citep{Ishimaru15,Ojima18}. This is supported by observations of the surviving ultra-faint dwarf galaxy Reticulum II. The metal-poor stars in Reticulum II formed from gas that was enriched by a prolific $r$-process event, believed to be a neutron star merger \citep{Ji16b}. 

In this paper, we investigate the possibility that metal-poor stellar halo stars with strong $r$-process signatures originated primarily in now-destroyed ultra-faint dwarf galaxies (UFDs) similar to Reticulum II. If this is true, the $r$-process stars we observe today could play a key role in understanding how the smallest building blocks of the Milky Way contribute to the \textbf{Galaxy's} formation.

This work is a first attempt to investigate this origin scenario of metal-poor $r$-process halo stars. We use cosmological models based on hierarchical galaxy formation simulations of Milky-Way-mass galaxies. 
This is described in Section \ref{sec:sim}, wherein we discuss our simulations and compare them to observed stellar halos (i.e., those of MW, M31, and GHOSTS galaxies). Star formation and chemical enrichment in low-mass galaxies is a field still under development (e.g., it is still difficult for \textbf{semi-analytic models} to reproduce even the mass-metallicity relation, \citealt{2017ApJ...846...66L}). We thus use empirical relations and parameterized models. In Section \ref{sec:rproc}, we describe our simple, empirically motivated treatment of $r$-process enrichment of early low-mass UFDs. Our results and a detailed discussion of the limitations of our model are found in Section \ref{sec:results}. Our conclusions are summarized in Section \ref{sec:conc}.

\section{Simulations} \label{sec:sim}

We analyze 31 dark matter only cosmological simulations of Milky-way-mass halos from the \emph{Caterpillar Project} \citep{Griffen16}. The zoom-in simulations in this suite have an effective resolution of $16{,}384^3$ particles of mass $3 \times 10^4$ M\textsubscript{$\odot$} in and around the galaxies of interest, resolving halos down to total mass ${\sim}10^6$ M\textsubscript{$\odot$}. The temporal resolution is ∼5 Myrs/snapshot from z = 31 to z = 6 and ∼50 Myrs to $z = 0$. 
The simulated halos in the suite span an unbisaed range of accretion histories.
For our analysis, we selected the simulated halos that were most Milky-Way-like, removing the halos that experienced late major mergers.

We briefly summarize details of how the simulations were developed \citep[for a more extensive explanation, see][]{Griffen16}. The halos in the zoom-in simulations were selected from a larger, lower resolution parent simulation in which structure evolved in a periodic box of comoving length 100 $h^{-1}$ Mpc with $1{,}024^3$ particles of mass $1.22 \times 10^7$ M\textsubscript{$\odot$}. The cosmological parameters were adopted from Planck 2013 $\Lambda$CDM cosmology: $\Omega_m = 0.32$, $\Omega_\Lambda = 0.68$, $\Omega_b = 0.05$, $\sigma_8 = 0.83$, $n_s = 0.96$, and H = 100 $h$ km s\textsuperscript{-1} Mpc\textsuperscript{-1} = 67.11 km s\textsuperscript{-1} Mpc\textsuperscript{-1} \citep{planck14}. Initial conditions were constructed using \texttt{MUSIC} \citep{music}. In the zoom-in simulations, care was taken to ensure that only the high-resolution volume of the Milky Way at $z > 10$ is studied and that no halos are contaminated with low-resolution particles. 
Dark matter subhalos were identified using a modified version of \texttt{ROCKSTAR} \citep{rockstar,Griffen16} and mergers trees were constructed by \texttt{CONSISTENT-TREES} \citep{consistenttrees}. The halos were assigned a virial mass $M_{vir}$ and radius $R_{vir}$ using the evolution of the virial relation from \citet{Bryan98}. For our cosmology, this corresponds to an overdensity of $\Delta_{crit} = 104$ at $z=0$.

To define the ``main branch" and ``destroyed subhalos" of a final $z=0$ halo (called the ``host halo"), we trace back the progenitors of the host halo at each simulation time step. At a given time step, the most massive progenitor of the host halo is a member of the ``main branch" and all other direct progenitors that merge into main branch halos are the ``destroyed subhalos". A subhalo is considered destroyed when it is no longer found by the halo finder.

\subsection{Assigning Stellar Mass and Metallicity to Subhalos}

Since the \emph{Caterpillar} halos only include a dark matter component, we incorporate luminous material through empirical relations, following \citet{Deason16}. 
For the results shown in this paper, we use the $M_{star}-M_{peak}$ relation derived by \citet{GK17} to estimate the stellar mass in each destroyed subhalo. 
$M_{peak}$ is defined as the peak virial mass from a subhalo's history. We also test the $M_{star}-M_{peak}$ relations derived by \citet{GK14}, \citet{Brook14}, \citet{Moster13}, and \citet{Behroozi13} \citep[see][]{Dooley17a}. The effects of the different relations are discussed in Section \ref{subsec:lim}. 

We also use empirical relations to estimate the metallicity of the stellar mass in the destroyed subhalos. We adopt a mass-metallicity ($M_{star}-\langle\mbox{[Fe/H]}\rangle$)\footnote{For elements A and B, $\mbox{[A/B]} \equiv \log(N_{A}/N_{B}) - \log(N_{A}/N_{B})_\odot $, where $N_{A}$ represents the abundance of $A$.} relation based on the $z=0$ relation determined by \citet{Kirby13b} for dwarf galaxies:
\begin{equation}
\langle\mbox{[Fe/H]}\rangle = (-1.69 \pm 0.04) + (0.30 \pm 0.02) \log \left(\frac{M_{star}}{10^6 M_{\odot}} \right)
\end{equation}

This $z=0$ relation is combined with the redshift evolution found by \citet{2016MNRAS.456.2140M} from hydrodynamical simulations: $\Delta \mbox{[Fe/H]} = 0.67 [e^{-0.5 z} - 1]$. This redshift evolution is consistent with observations \citep{Leethochawalit18}. For the destroyed subhalos that are sufficiently massive to form stars after reionization, we use the redshift of their destruction ($z_{dest}$) as the redshift at which to determine their mean metallicity. Determining their metallicity at other redshifts (e.g., the redshift at which they reach peak mass, $z_{peak}$, or the redshift at first infall, $z_{infall}$) does not significantly affect results. For subhalos that form stars before reionization but have their star formation permanently suppressed (e.g., UFDs; for our full definition of UFDs see Section \ref{subsec:UFDdef}), we use $z=0$ as the redshift at which to determine their mean metallicity. This is because the UFDs observed today at $z=0$ also stopped forming stars long ago and thus will appear similar (at least in metallicity) to the UFDs that were destroyed.

After determining the mean metallicity of each subhalo, we assume a Gaussian distribution about the mean with standard deviation of 0.4 dex. This standard deviation aligns with the observed intrinsic scatter for dwarfs at $z=0$ \citep{Deason16}. The metallicity distribution function (MDF) of each individual destroyed subhalo is weighted by the stellar mass of the halo and combined to form the MDF of the accreted portion of the stellar halo. Our resolution supports metallicities down to about $\mbox{[Fe/H]} \sim -4.5$; below this metallicity, the MDF receives a greater than 1\% contribution from unresolved halos.

These methods of assigning stellar mass and metallicity to the destroyed subhalos are nearly the same methods used by \citet{Deason16}. The two significant differences are (1) our use of an updated $M_{star}-M_{peak}$ relation that assumes increased scatter about the relation for lower mass halos and (2) our use of $z=0$ as the redshift at which to determine the metallicity of destroyed UFDs. While they used $z_{dest}$ instead of $z=0$ for UFD metallicity, they acknowledge that $z=0$ is likely the appropriate redshift to use. They only did not use $z=0$ because the metallicity of the UFDs did not affect the bulk properties they were interested in.

\subsection{Mass Scales for Star Formation} \label{subsec:massscales}

\begin{table*}[t!]
\centering
\caption{Values For Model Parameters} \label{tab:mass}
\begin{tabular}{c|c|cc}
\tablewidth{0pt}
\hline
\hline
 & Fiducial & Values & Justification \\
\hline
$z_{reion}$ & 8 & 6, 8, 10, 12 & From radiation-hydrodynamic simulation\textsuperscript{1}, $\langle z_{reion} \rangle = 7.8$ \\
\hline
$M_{SF}$ & $5 \times 10^7$ $M_{\odot}$ & $10^8$ $M_{\odot}$ & Atomic cooling threshold\textsuperscript{2}, corresponds to $T_{vir} \sim 10^4$ K \\
 & & $5 \times 10^7$ $M_{\odot}$ & Results in ${\sim}120$ surviving UFDs\textsuperscript{3} when $z_{reion}=8$ \\
\hline
$M_{filt}$ & $2 \times 10^9$ $M_{\odot}$ & $6 \times 10^9$ $M_{\odot}$ & From hydrodynamical simulations\textsuperscript{4}, corresponds to $v_{max} \sim 25$ km/s  \\
 & & $2 \times 10^9$ $M_{\odot}$ & From radiation-hydrodynamic simulations of reionization\textsuperscript{5} \\
\hline
$f_{NSM}$ & 10\% & $5-15$\% & Percentage of UFDs observed to be $r$-process enhanced\textsuperscript{6} \\
\hline
$M_{UFD,max}$ & $10^9$ $M_{\odot}$ & varied & Max mass for a halo to be highly enriched after a single $r$-process event\textsuperscript{7} \\
\hline
\end{tabular}
\\ \textsuperscript{1}\citet{2018ApJ...856L..22A}; \textsuperscript{2}\citet{Bromm11}; \textsuperscript{3}see Section \ref{subsec:massscales} and \citet{2018MNRAS.479.2853N}; 
\\  \textsuperscript{4}\citet{2008MNRAS.390..920O}; \textsuperscript{5}\citet{2016MNRAS.463.1462O}; \textsuperscript{6}see Section \ref{subsec:fNSM}; \textsuperscript{7}see Section \ref{subsec:UFDdef}
\end{table*}

To determine which destroyed subhalos have their star formation permanently suppressed by reionization, which subhalos restart star formation after reionization, and which subhalos never form stars, we adopt cutoffs at different halo mass scales \citep[e.g.,][]{Dooley17a}. These mass scales, $M_{SF}$ and $M_{filt}$, are summarized in Table \ref{tab:mass}.

We also assume instantaneous reionization. The choice of reionization redshift is most important for the stellar mass of low-mass halos. Using a radiation-hydrodynamics simulation of Milky-Way-like galaxies, \citet{2018ApJ...856L..22A}  found that progenitor halos with $M_{vir}(z=0) < 10^{11}$ $M_\odot$ reionized around the globally averaged 50\% reionization at $\langle z_{reion} \rangle = 7.8$. We therefore assume $z_{reion} \sim 8$, but investigate several possible reionization redshifts ($z_{reion}=6,8,10,12$).

$M_{SF}$ is the minimum halo mass needed to form stars. One option for $M_{SF}$ is $10^8$ $M_{\odot}$, corresponding to $T_{vir} \sim 10^4$ K. This is motivated by the atomic cooling threshold a halo must exceed before star formation can be efficiently sustained \citep{Bromm11}. We also investigate a slightly lower choice for $M_{SF}$. Our choice of $M_{SF}$ significantly changes the number of surviving satellite UFDs at $z=0$. $M_{SF} = 10^8$ $M_{\odot}$ results in only ${\sim}40$ surviving UFDs. About 40 surviving UFDs around the Milky Way have already been discovered and many more are expected to be found, so this number is low \citep{Dooley17a}. \citet{Graus18} also recently found that the threshold for $M_{SF}$ must be lowered to match the observed number of satellites. We therefore adopt $M_{SF} = 5 \times 10^7$ $M_{\odot}$. This choice results in each simulation having ${\sim}120$ surviving UFDs at $z=0$ (assuming $z_{reion}=8$), which is roughly the number expected to exist around the Milky Way \citep{2018MNRAS.479.2853N}. We also tested $M_{SF} = 7 \times 10^7$ $M_{\odot}$, but this choice results in each simulation having ${\sim}70$ surviving UFDs at $z=0$, which is too few.

$M_{filt}$ is the filtering mass, the mass below which galaxies are significantly affected by the photoionizing background. A halo must surpass this mass scale to continue star formation after reionization \citep{Gnedin00}. Using hydrodynamical simulations of low mass halos in an ionizing background, \citet{2008MNRAS.390..920O} found that halos with circular velocities below ${\sim}25$ km s\textsuperscript{-1} (corresponding to $M_{filt} \sim 6 \times 10^9$ $M_\odot$) lose a significant amount of their gas due to photoheating. More recent radiation-hydrodynamic simulations of reionization by \citet{2016MNRAS.463.1462O} find that photoheating suppresses the star formation of halos below $M_{filt} \sim 2 \times 10^9$ $M_{\odot}$. The filtering mass scale is still uncertain, so we try both of these thresholds.

Of the around 20,000 resolved subhalos that are destroyed into each of our 31 host halos, fewer than 100 subhalos become massive enough to ever form stars. If a subhalo has $M_{vir} < M_{SF}$ at reionization and $M_{peak} < M_{filt}$, it does not form stars prior to reionization and has its star formation permanently suppressed by reionization, meaning it does not ultimately contribute to a stellar halo. Subhalos that form after reionization but remain low mass (never surpassing the mass threshold for star formation; $M_{peak} > M_{SF}$) also do not form stars.

These mass scales only affect low-mass halos. Because the stellar halo is dominated by only a few high-mass destroyed halos, these mass scales do not affect bulk properties of the stellar halo. They are significant for the low-metallicity portion of the stellar halo that we are interested in, however.

  
\subsection{Simulated Stellar Halos vs. Observed Stellar Halos} \label{subsec:compare}

To verify that our stellar mass and metallicity estimations are reasonable, we compare the properties of the simulated stellar halos to those of real, observed stellar halos. Despite the similarities between the \textit{Caterpillar} stellar halos and the observed stellar halos, however, we caution that our simulated stellar halos are formed exclusively from accreted stars while actual stellar halos are not. Some accreted stars inevitably end up in the disk/bulge and some \textit{in situ} disk/bulge stars inevitably end up being thrown into the stellar halo \citep[e.g.,][]{Cooper15,Gomez17}. Thus while we will use the terms ``stellar halo" and ``accreted stars" somewhat interchangeably, there is a difference. Our simulated stellar halos are approximations of actual stellar halos. We discuss the effects of this approximation in Section \ref{subsec:lim}.

\begin{figure}[t!]
\epsscale{1.15}
\plotone{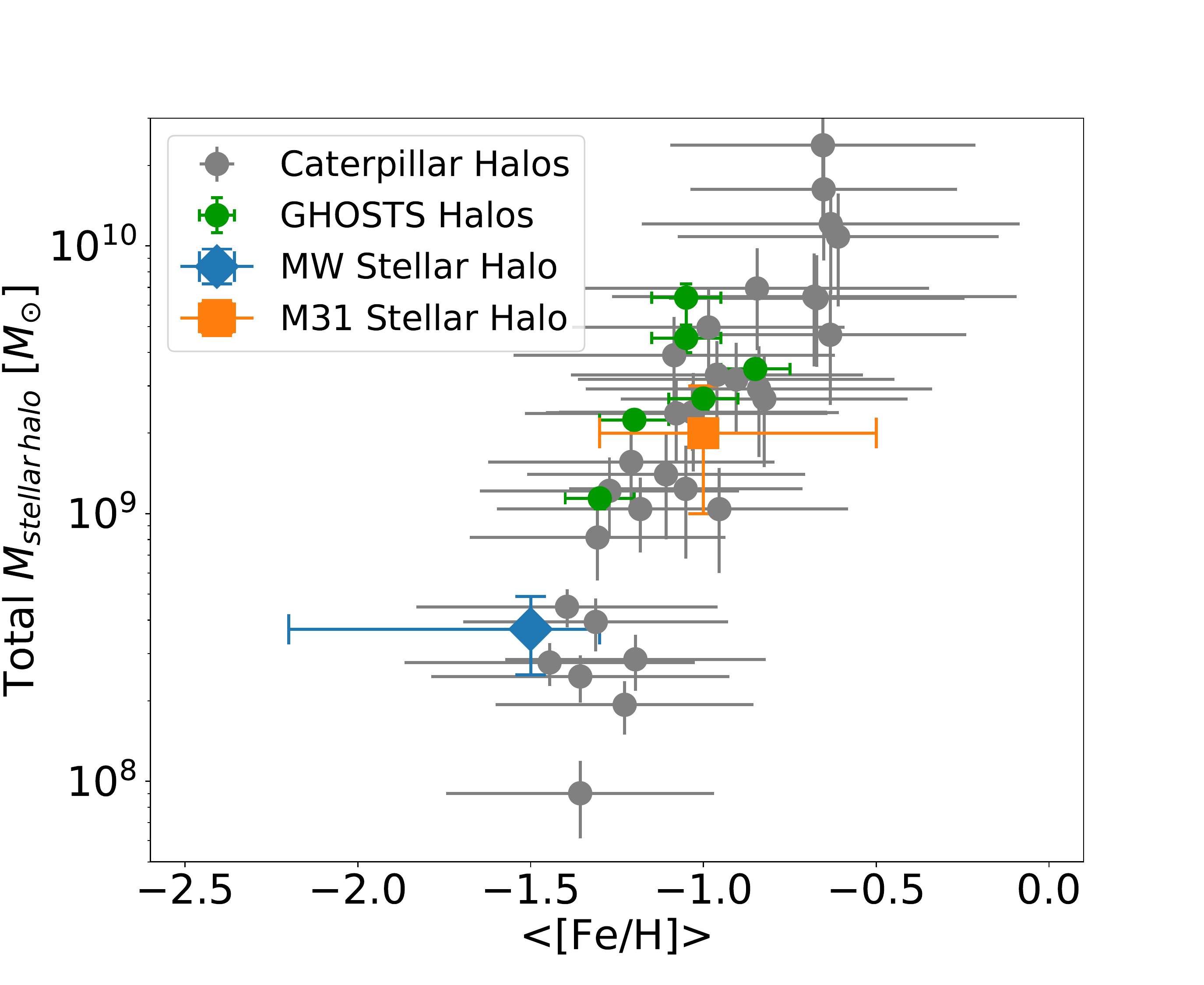}
\caption{The average metallicity and total stellar mass of the \textit{Caterpillar} stellar halos compared to the stellar halos of the Milky Way, M31, and galaxies in the GHOSTS survey. The span of the \textit{Caterpillar} stellar halos well captures the bulk properties of these galaxies and their relative differences. \label{fig:ZvsM}}
\end{figure}

\begin{figure*}
\gridline{\fig{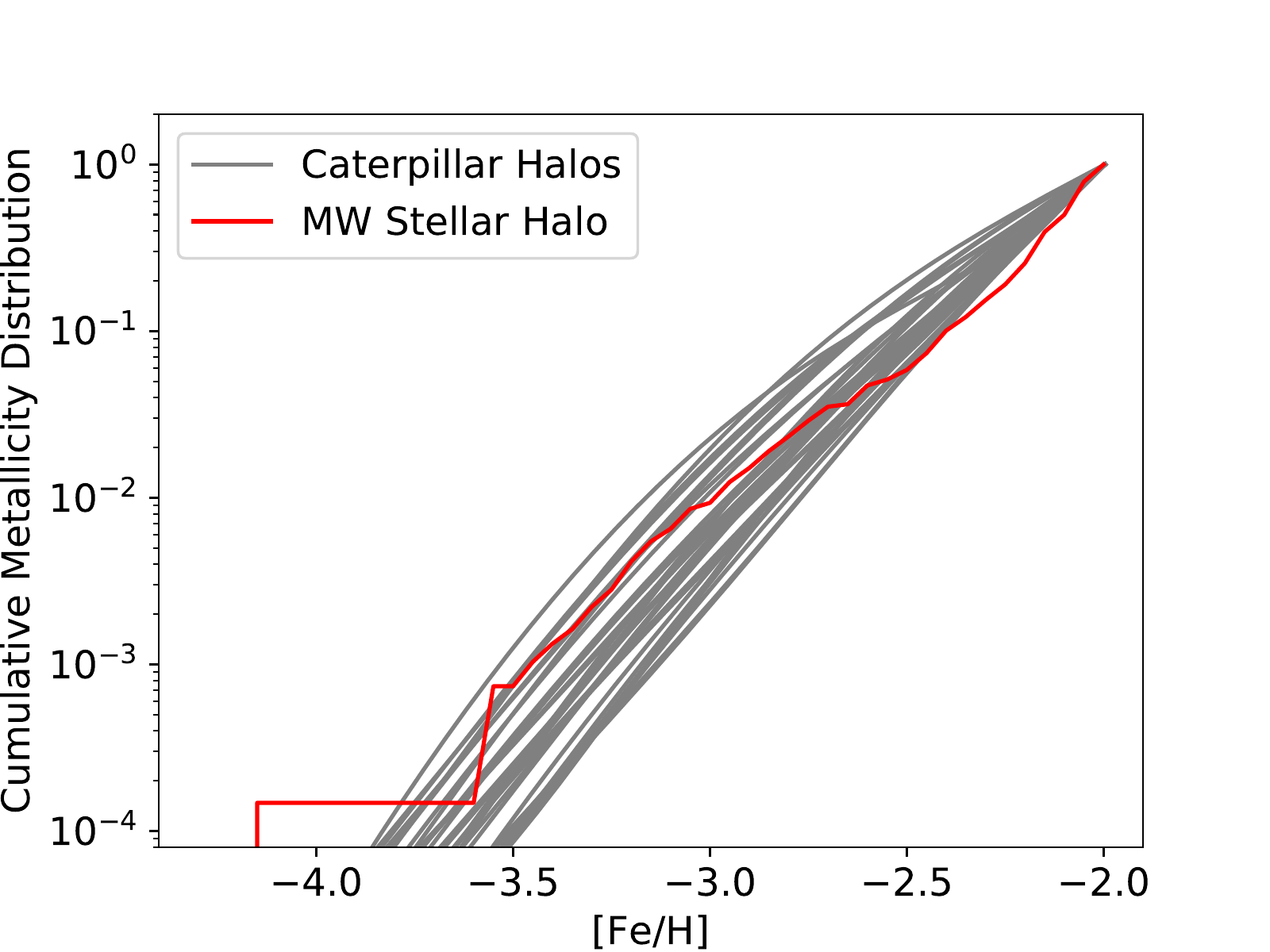}{0.5\textwidth}{(a) $M_{SF} = 5 \times 10^7$ $M_{\odot}$, $M_{filt} = 2 \times 10^9$ $M_{\odot}$}
		  \fig{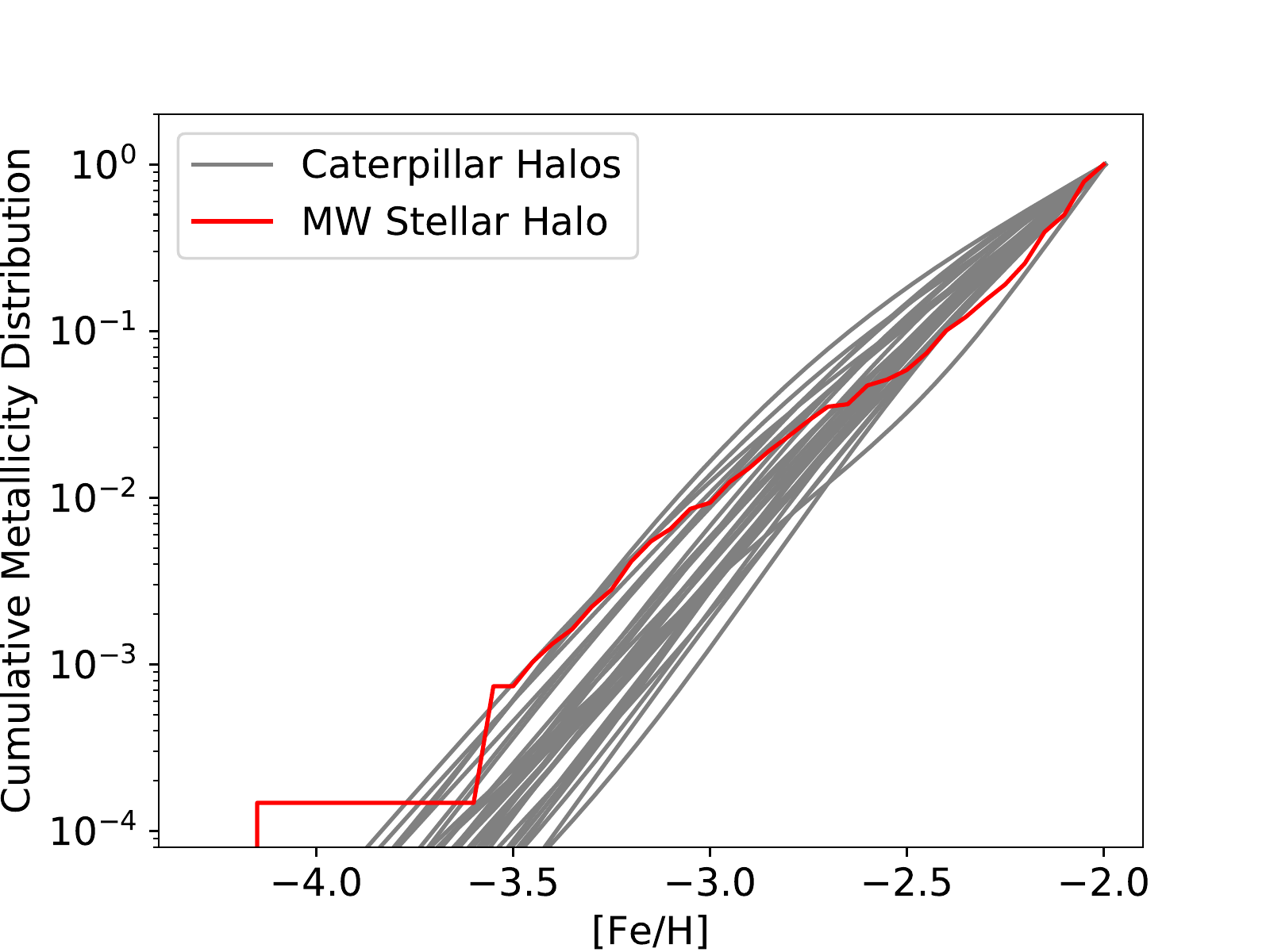}{0.5\textwidth}{(b) $M_{SF} = 5 \times 10^7$ $M_{\odot}$, $M_{filt} = 6 \times 10^9$ $M_{\odot}$}}
\caption{Cumulative metallicity distribution functions for stellar halo stars with $\mbox{[Fe/H]} < -2$. The \textit{Caterpillar} halos are compared to the Milky Way stellar halo \citep{2009A&A...507..817S}. Results do not noticeably change for $M_{SF}=10^8$ $M_{\odot}$. Assumes $z_{reion} = 8$.
\label{fig:CDF}}
\end{figure*}

Figure \ref{fig:ZvsM} compares the average metallicity and total stellar mass of each of the \textit{Caterpillar} stellar halos (composed entirely of \textit{ex situ} stars) to those of observed stellar halos. We compare to galaxies in the GHOSTS survey \citep[NGC253, NGC891, M81, NGC4565, NGC4945, and NGC7814;][]{Monachesi16,Harmsen17}, the Milky Way ($M_{star,halo} \sim 3.7 \pm 1.2 \times 10^8$ $M_{\odot}$, \citealt{Bell08}; $\langle\mbox{[Fe/H]}\rangle \sim -1.3 \text{ to } -2.2$, \citealt{2010ApJ...712..692C}), and M31 ($M_{star,halo} \sim 2 \pm 1 \times 10^9$ $M_{\odot}$, \citet{2015ApJ...802...49W}; $\langle\mbox{[Fe/H]}\rangle \sim -0.5 \text{ to } -1.3$, \citet{2006ApJ...648..389K}). Given the simplicity of the model, the span of the \textit{Caterpillar} stellar halos matches the properties of these observed halos and their relative differences remarkably well. Figure \ref{fig:ZvsM} is a recreation of Figure 8 from \citet{Deason16}, which also reproduces the relative difference between the Milky Way and M31. 

We also compare the cumulative metallicity distribution function of the \textit{Caterpillar} stellar halos to that of the Milky Way. In Figure \ref{fig:CDF} we compare to the cumulative distribution function from \citet{2009A&A...507..817S} for metal-poor halo stars with metallicity $\mbox{[Fe/H]}<-2$. Below $\mbox{[Fe/H]}<-3.5$, the \textit{Caterpillar} distributions differ significantly from the Milky Way distribution. This is likely because the Gaussians composing the MDFs have a weaker metal-poor tail than the actual distributions in each destroyed subhalo. However, the composite Gaussian MDFs provide a much better fit than other physically motivated MDFs (e.g., Extra Gas model, power laws), so we keep the Gaussian MDFs for this analysis. We discuss the limitations of these empirical relations and fixed [Fe/H] distributions in Section \ref{subsec:lim}.

\section{Treatment of $r$-Process Enrichment in Early UFDs} \label{sec:rproc}

We assume that some fraction (see Section \ref{subsec:fNSM}) of now-destroyed UFDs experience an early neutron star merger (NSM) or another rare prolific $r$-process event that enriches the gas from which subsequent stars form. Looking specifically at low-mass UFDs, i.e. dwarf galaxies small enough to form highly $r$-process enhanced stars after only one $r$-process event, we consider the stars formed in these  galaxies to be r-II stars. r-II stars are stars that are highly enhanced in $r$-process elements: $\mbox{[Eu/Fe]} > +1$, $\mbox{[Ba/Eu]} < 0$ \citep{Beers05}. This follows the example set by Reticulum II. We note that the mass range of UFDs is not universally defined, but when creating our definition of UFDs we focus on the low-mass end (see Section \ref{subsec:UFDdef}) to look specifically at the smallest galaxies that helped form the Milky Way.

In each simulation, we then trace all the galaxies that disrupt into each host galaxy to $z=0$ and compare the fraction of simulated r-II stars from destroyed UFDs ($f_{r-II,sim}$; see Section \ref{subsec:frII}) to the observed fraction of r-II stars in the Milky Way's stellar halo ($f_{r-II,obs}$). In this way, we investigate how much of the observed fraction of r-II halo stars may have originated in destroyed ultra-faint dwarfs.

This treatment only considers putative r-II stars that form from gas enriched by a single $r$-process event in a low-mass destroyed galaxy. Actual r-II halo stars can also form through other pathways, e.g., a higher-mass destroyed galaxy that experiences more than one NSM event could form r-II stars, or r-II stars could form \emph{in situ} through inhomogeneous mixing and later get thrown into the stellar halo \citep{Shen15,vanDeVoort15,Naiman18}. To specifically investigate the origins of r-II halo stars in low-mass UFDs, though, we do not simulate r-II stars with other origins. We discuss the limitations of this analysis in more depth in Section \ref{subsec:lim}.

\subsection{Definition of Ultra-Faint Dwarf} \label{subsec:UFDdef}

We define an ultra-faint dwarf as a halo that forms stars early in the Universe's history ($M_{vir} > M_{SF}$ before reionization), but has its star formation permanently suppressed by reionization ($M_{peak} < M_{filt}$). This is the ``fossil" definition of UFDs \citep[e.g.,][]{Bovill11}. We also require that UFDs have a final $M_{star} < 2 \times 10^5$ $M_\odot$ (corresponding to $M_{peak} \lesssim 2.8 \times 10^9$ $M_\odot$\footnote{Using the $M_{star}-M_{peak}$ relation derived by \citet{GK17}}). When identifying now-destroyed UFDs (the smallest building blocks of the galaxy), we consider both UFDs that disrupted directly into the main branch of the host halo and UFDs that disrupted into other dwarf galaxies before merging with the host halo.

Furthermore, we constrain our definition of UFDs to only include halos in which a single prolific $r$-process event can enrich the gas to produce subsequent r-II stars as defined by high [Eu/Fe]. This excludes ``high mass" UFDs because of dilution.
We define $M_{UFD,max}$ as the maximum mass a UFD can reach while it is forming stars (before reionization). More massive subhalos would dilute the chemical enrichment products too much to still yield r-II stars after a single NSM event. The calculations to determine $M_{UFD,max}$ are uncertain, however.

For example, in Reticulum~II, the prolific $r$-process event resulted in stars with $\mbox{[Eu/H]} \sim -1.3$ \citep{Ji16b}. This corresponds to ${\sim}10^{-4.5}$ $M_\odot$ of Eu being injected into ${\sim}10^6$ $M_\odot$ of gas (possibly an order of magnitude higher or lower; in a ${\sim}10^{7-8}$ $M_\odot$ halo). Such an event produces r-II stars at $\mbox{[Fe/H]} \lesssim -2.3$. Using proportionality arguments, a halo's mixing mass can be related to its virial mass: $M_{mix} \sim M_{vir}^{1.25}$ \citep[e.g.,][]{Ji15}. Increasing the virial mass by an order of magnitude would increase the mixing mass by $10^{1.25}$, producing r-II stars at lower metallicities: around $\mbox{[Fe/H]} \lesssim -2.3 -1.25 = -3.55$. It is therefore unreasonable to assume all of the stellar mass below $\mbox{[Fe/H]} < -2.5$ in ``high mass" UFDs would be highly $r$-process enhanced following one NSM event.

We caution that the mixing mass numbers are highly uncertain and based on order-of-magnitude arguments. We thus try several different maximum mass cutoffs: $M_{UFD,max}=2 \times 10^8$, $5 \times 10^8$, $10^9$, and $2 \times 10^9$ $M_\odot$. Of these, $2 \times 10^8$ in particular is quite low because the minimum mass of a UFD is $M_{SF}\sim 10^8$ $M_\odot$, but we include it to encompass the possible parameter values. We use the intermediate choice of $10^9$ $M_\odot$ for our fiducial model.



\subsection{Neutron Star Merger Fraction, $f_{NSM}$} \label{subsec:fNSM}

\textbf{We empirically determine the fraction of UFDs that experienced r-process enhancement simply by comparing the number of known r-process UFDs to normal UFDs.} There are now high-resolution spectroscopic abundances for stars in 15 surviving UFDs: Bootes~I \citep{Feltzing09,Frebel16}, Bootes~II \citep{Koch14b,Ji16a}, Coma Berenices~II \citep{Francois16}, Coma Berenices \citep{Frebel10b}, Grus~I \citep{Ji18prep}, Hercules \citep{Koch13}, Horologium~I \citep{Nagasawa18}, Leo IV \citep{Simon10}, Reticulum~II \citep{Ji16b}, Segue 1 \citep{Frebel14}, Segue 2 \citep{Roederer14a}, Triangulum II \citep{Venn17,Kirby17}, Tucana II \citep{Ji16d}, Tucana III \citep{Hansen17}, and Ursa Major II \citep{Frebel10b}.

Of these, Reticulum~II has definitely been enriched by a prolific $r$-process event, assumed to be a neutron star merger \citep{Ji16b}.
Tucana~III also exhibits $r$-process enhancement \citep{Hansen17}, though it is still unclear if this is a tidally disrupted UFD or a globular cluster \citep{Li18}.
It thus seems that there are $1-2$ UFDs affected by an $r$-process event out of $14-15$ UFDs, or 7.1\% to 13.3\%.

For the purposes of this analysis, we therefore assume $5-15$\% of now-destroyed UFDs experience an early neutron star merger (NSM) or some other rare prolific $r$-process event. Our default NSM fraction is $f_{NSM}\sim 10$\%. This fraction is agnostic to the actual nature of the $r$-process event; it directly relates to the fraction of surviving UFDs that have been observed to be $r$-process enhanced.

\subsection{r-II Star Fraction, $f_{r-II}$} \label{subsec:frII}

The simulated r-II star fraction, $f_{r-II,sim}$, is the amount of low-metallicity, highly $r$-process enhanced stars that we assume originated in now-destroyed low-mass UFDs compared to all low-metallicity stars now present in the accreted stellar halo.

\begin{equation}
f_{r-II,sim} = \frac{\text{metal-poor r-II halo stars that formed in UFDs}} {\text{all metal-poor halo stars}}
\end{equation}

We define ``metal-poor'' as $\mbox{[Fe/H]} < -2.5$. For the simulated fraction, the numerator and denominator are both in units of stellar mass as opposed to numbers of stars. Our methodology directly estimates the amount of stellar mass in each galaxy, not the number of stars in each galaxy, but this makes little difference for old stellar populations. To determine how much of the stellar mass in a galaxy is metal-poor, we integrate the MDF below $\mbox{[Fe/H]} = -2.5$.

The ``as observed" r-II star fraction, $f_{r-II,obs}$, includes all currently known r-II stars in the Milky Way's stellar halo.

\begin{equation}
f_{r-II,obs} = \frac{\text{metal-poor r-II halo stars}} {\text{all metal-poor halo stars}}
\end{equation}
Observed r-II stars are stars which display strong $r$-process enhancement ($\mbox{[Eu/Fe]} > 1$ and $\mbox{[Ba/Eu]} < 0$). For the observed fraction, the numerator and denominator are both in terms of the number of observed stars.

\section{Results and Discussion} \label{sec:results}    

\subsection{Simulated Fraction of r-II Stars in the Stellar Halo} \label{subsec:fracs}

\begin{figure*}[t!]
\plotone{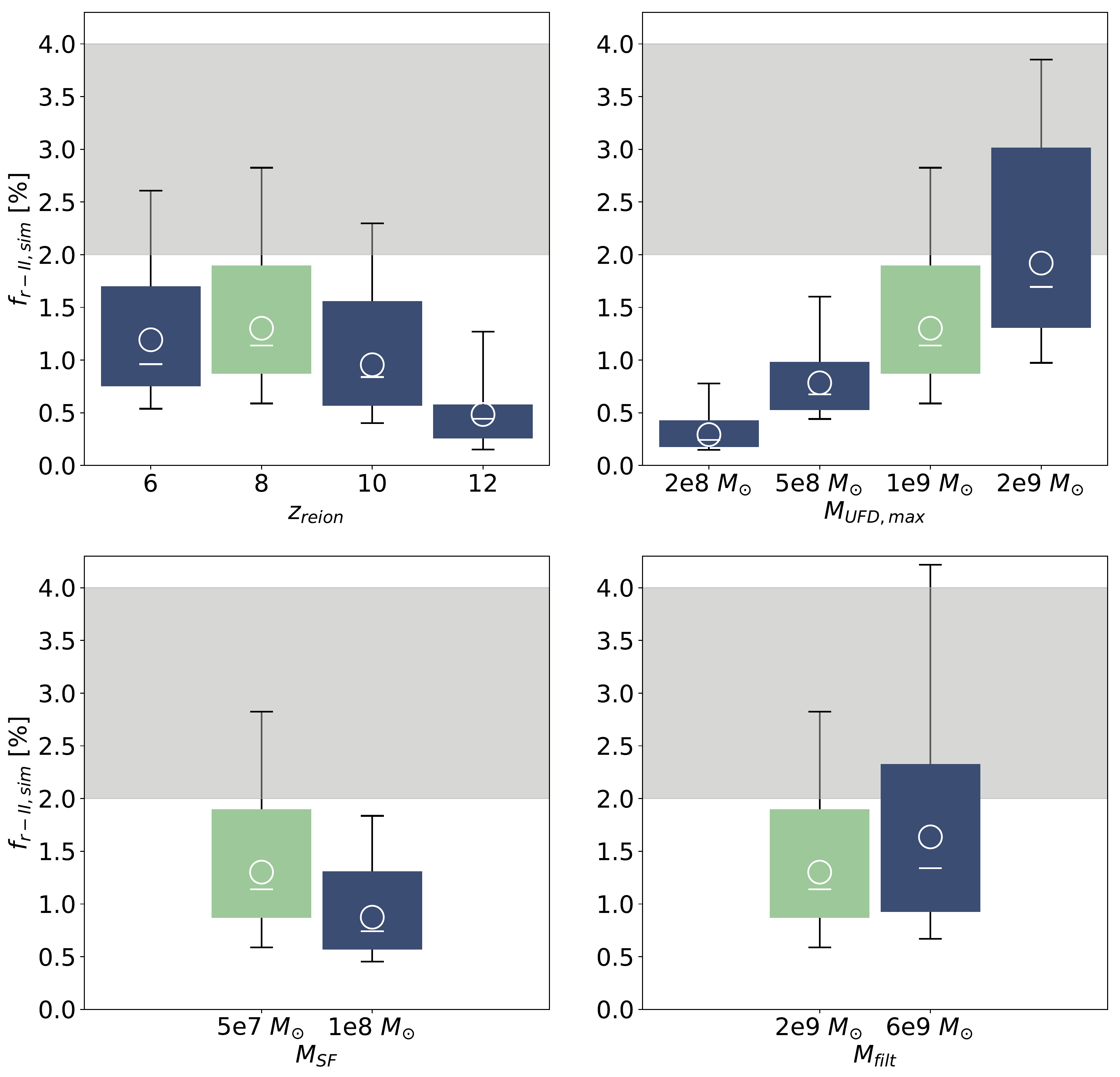}
\caption{The simulated r-II star fraction, $f_{r-II,sim}$, as it varies with different mass thresholds and reionization redshifts. For each set of parameters, the mean $f_{r-II,sim}$ is shown as a white circle and the median is shown as a white line. The colored boxes correspond to 68\% scatter between simulations, and the error bars shown the minimum and maximum $f_{r-II,sim}$. Our fiducial model is highlighted in light green, and single-parameter variations on the fiducial model are shown in blue. The currently observed fraction of r-II stars in the Milky Way stellar halo ($f_{r-II,obs} \sim 2-4$\%) is shown in grey. See Section \ref{subsec:frII} for definitions of the r-II fractions and Table \ref{tab:mass} for explanations of the different parameters.
\label{fig:fracs}}
\end{figure*}

\begin{table*}[t!]
\centering
\caption{The simulated r-II star fraction, $f_{r-II,sim}$, for different mass thresholds and reionization redshifts. The reported values are averaged across simulations and the uncertainties correspond to 68\% scatter. The fiducial model is bolded. Variations of $f_{NSM}$ are not explicitly shown because $f_{r-II,sim}$ scales linearly with $f_{NSM}$.} \label{tab:fracs}
\begin{tabular}{ccccc}
\tablewidth{0pt}
\hline
\hline
 & $M_{SF} = 5 \times 10^7$ $M_{\odot}$  & $M_{SF} = 5 \times 10^7$ $M_{\odot}$  & $M_{SF} = 10^8$ $M_{\odot}$    & $M_{SF} = 10^8$ $M_{\odot}$  \\
 & $M_{filt} = 2 \times 10^9$ $M_{\odot}$  & $M_{filt} = 6 \times 10^9$ $M_{\odot}$ & $M_{filt} = 2 \times 10^9$ $M_{\odot}$ & $M_{filt} = 6 \times 10^9$ $M_{\odot}$ \\
\hline
\multicolumn{5}{c}{$M_{UFD,max} = 2 \times 10^9$ $M_{\odot}$} \\
\hline
$z_{reion} = 6$ & $1.7^{+0.9}_{-0.5}$ \% & $2.4^{+0.8}_{-0.8}$ \%& $1.5^{+0.8}_{-0.7}$ \% & $2.1^{+0.8}_{-0.6}$ \%  \\
$z_{reion} = 8$ & $1.9^{+1.0}_{-0.6}$ \% & $2.7^{+0.9}_{-0.8}$ \% & $1.5^{+0.9}_{-0.6}$ \% & $2.2^{+0.8}_{-0.7}$ \% \\
$z_{reion} = 10$ & $1.5^{+0.8}_{-0.5}$ \% & $2.3^{+0.7}_{-0.7}$ \% & $1.0^{+0.3}_{-0.3}$ \% & $1.5^{+0.6}_{-0.4}$ \% \\
$z_{reion} = 12$ & $0.9^{+0.4}_{-0.2}$ \% & $1.6^{+0.4}_{-0.4}$ \% & $0.4^{+0.3}_{-0.2}$ \% & $0.8^{+0.2}_{-0.2}$ \% \\
\hline
\multicolumn{5}{c}{$M_{UFD,max} = 10^9$ $M_{\odot}$} \\
\hline
$z_{reion} = 6$ & $1.2^{+0.5}_{-0.4}$ \% & $1.6^{+0.6}_{-0.5}$ \% & $0.9^{+0.5}_{-0.3}$ \% & $1.4^{+0.5}_{-0.5}$ \% \\
$z_{reion} = 8$ & $\mathbf{1.3^{+0.6}_{-0.4}}$ \% & $1.6^{+0.7}_{-0.7}$ \% & $0.9^{+0.4}_{-0.3}$ \% & $1.5^{+0.5}_{-0.4}$ \% \\
$z_{reion} = 10$ & $1.0^{+0.6}_{-0.4}$ \% & $1.4^{+0.6}_{-0.5}$ \% & $0.5^{+0.2}_{-0.1}$ \% & $0.9^{+0.4}_{-0.3}$ \% \\
$z_{reion} = 12$ & $0.5^{+0.1}_{-0.2}$ \% & $0.9^{+0.4}_{-0.3}$ \% & $0.2^{+0.1}_{-0.1}$ \% & $0.4^{+0.2}_{-0.1}$ \% \\
\hline
\multicolumn{5}{c}{$M_{UFD,max} = 5 \times 10^8$ $M_{\odot}$} \\
\hline
$z_{reion} = 6$ & $0.7^{+0.2}_{-0.2}$ \% & $1.1^{+0.4}_{-0.3}$ \% & $0.5^{+0.2}_{-0.2}$ \% & $0.9^{+0.4}_{-0.2}$ \% \\
$z_{reion} = 8$ & $0.8^{+0.2}_{-0.3}$ \% & $1.3^{+0.4}_{-0.4}$ \% & $0.5^{+0.2}_{-0.1}$ \% & $0.9^{+0.3}_{-0.3}$ \% \\
$z_{reion} = 10$ & $0.5^{+0.2}_{-0.2}$ \% & $1.0^{+0.3}_{-0.3}$ \% & $0.2^{+0.1}_{-0.1}$ \% & $0.5^{+0.2}_{-0.1}$ \% \\
$z_{reion} = 12$ & $0.2^{+0.1}_{-0.1}$ \% & $0.4^{+0.2}_{-0.1}$ \% & $0.1^{+0.1}_{-0.1}$ \% & $0.2^{+0.1}_{-0.1}$ \% \\
\hline
\multicolumn{5}{c}{$M_{UFD,max} = 2 \times 10^8$ $M_{\odot}$} \\
\hline
$z_{reion} = 6$ & $0.33^{+0.12}_{-0.10}$ \% & $0.47^{+0.25}_{-0.19}$ \% & $0.15^{+0.06}_{-0.05}$ \% & $0.22^{+0.11}_{-0.07}$ \% \\
$z_{reion} = 8$ & $0.29^{+0.14}_{-0.12}$ \% & $0.52^{+0.20}_{-0.12}$ \% & $0.11^{+0.03}_{-0.03}$ \% & $0.24^{+0.10}_{-0.07}$ \% \\
$z_{reion} = 10$ & $0.16^{+0.04}_{-0.06}$ \% & $0.32^{+0.05}_{-0.05}$ \% & $0.05^{+0.02}_{-0.03}$ \% & $0.09^{+0.06}_{-0.04}$ \% \\
$z_{reion} = 12$ & $0.07^{+0.03}_{-0.03}$ \% & $0.13^{+0.04}_{-0.04}$ \% & $0.02^{+0.01}_{-0.01}$ \% & $0.02^{+0.02}_{-0.01}$ \% \\
\hline
\end{tabular}
\end{table*}

Using the treatment of $r$-process enrichment described in the previous section, we calculate the simulated r-II star fraction, $f_{r-II,sim}$, for different $z_{reion}$ and mass thresholds. Figure \ref{fig:fracs} and Table \ref{tab:fracs} show these results. 

The simulated r-II fraction is ${\sim}1.3$\% for our fiducial parameter values ($z_{reion} = 8$, $M_{UFD,max} = 10^9$ $M_{\odot}$, $M_{SF} = 5 \times 10^7$ $M_{\odot}$, $M_{filt} = 2 \times 10^9$ $M_{\odot}$, $f_{NSM} = 10$\%). The fraction varies somewhat with all the parameters, as seen in Figure \ref{fig:fracs}. It scales linearly with the NSM fraction, $f_{NSM}$. If we consider $f_{NSM} = 5 - 15$\% with our fiducial model, the r-II fraction is ${\sim}0.7 - 2$\%. If we consider the scatter between the simulations, the r-II fraction is ${\sim}1 - 2$\%. Varying all of the parameters to the determine the minimum and maximum simulated r-II fraction gives a range of ${\sim}0.01 - 4$\% with the favored value being around $1-2$\%.

For comparison, the observed fraction, $f_{r-II,obs}$ differs a bit from sample to sample (e.g., 3.3\%, \citealt{2015ApJ...807..171J}; 2.2\%, \citealt{2014AJ....147..136R}; 2.9\%, \citealt{2005A&A...439..129B}). We also note that these fractions depend on the specifically chosen limit $\mbox{[Eu/Fe]} > 1$ and general sample selections that are not completeness corrected. A recent study from \citet{Hansen18} found $f_{r-II,obs} \sim 10$\%, but a larger sample has reduced the fraction by about half and data is still being collected (T. Hansen, priv. comm.). This study was also specifically looking for $r$-process stars and may not be representative of the true r-II fraction. Aggregating the surveys and individual reports in the literature without attempting to account for observational bias gives 3.2\% \citep{jinabase,Hansen18}. We note that r-II stars are preferentially likely to be reported in literature over other metal-poor stars, however, so r-II stars are probably overrepresented. Currently, $f_{r-II,obs}$  appears to be ${\sim}2-4$\%, but the true fraction is likely lower. 

Comparing the simulated $f_{r-II,sim} \sim 1-2$\% and the observed $f_{r-II,obs} \sim 2-4$\%, around half of the low-metallicity r-II halo stars could have originated in now-destroyed UFDs following a single $r$-process event. Considering the 68\% scatter between simulations and the effects of varying $M_{SF}$ and $M_{filt}$ in the fiducial model, $f_{r-II,sim} \sim 0.6-2.3$ can account for ${\sim}20-80$\% of the current $f_{r-II,obs}$. Varying $M_{UFD,max}$ from $5 \times 10^8$ $M_\odot$ to $2 \times 10^9$ $M_\odot$ expands the range to ${\sim}20-100$\% of the current ``as observed'' $f_{r-II,obs}$. Furthermore, because the true $f_{r-II,obs}$ is likely lower, the amount of the true $f_{r-II,obs}$ that $f_{r-II,sim}$ can account for is likely closer to ${\sim}80$\% than $20$\%. This implies that a significant fraction of the metal-poor r-II halo stars likely originated in now-destroyed UFDs. This is only considering the contribution of ``low-mass'' UFDs. Including other r-II star creation pathways (e.g., more than one NSM or inhomogeneous mixing in higher-mass UFDs) would increase the fraction. The caveats to this result are discussed in Section \ref{subsec:lim}. This result is supported by recent kinematic evidence that implies r-II stars were largely accreted. For more on this, see Section \ref{sec:conc}.

\begin{figure}[t!]
\epsscale{1.1}
\plotone{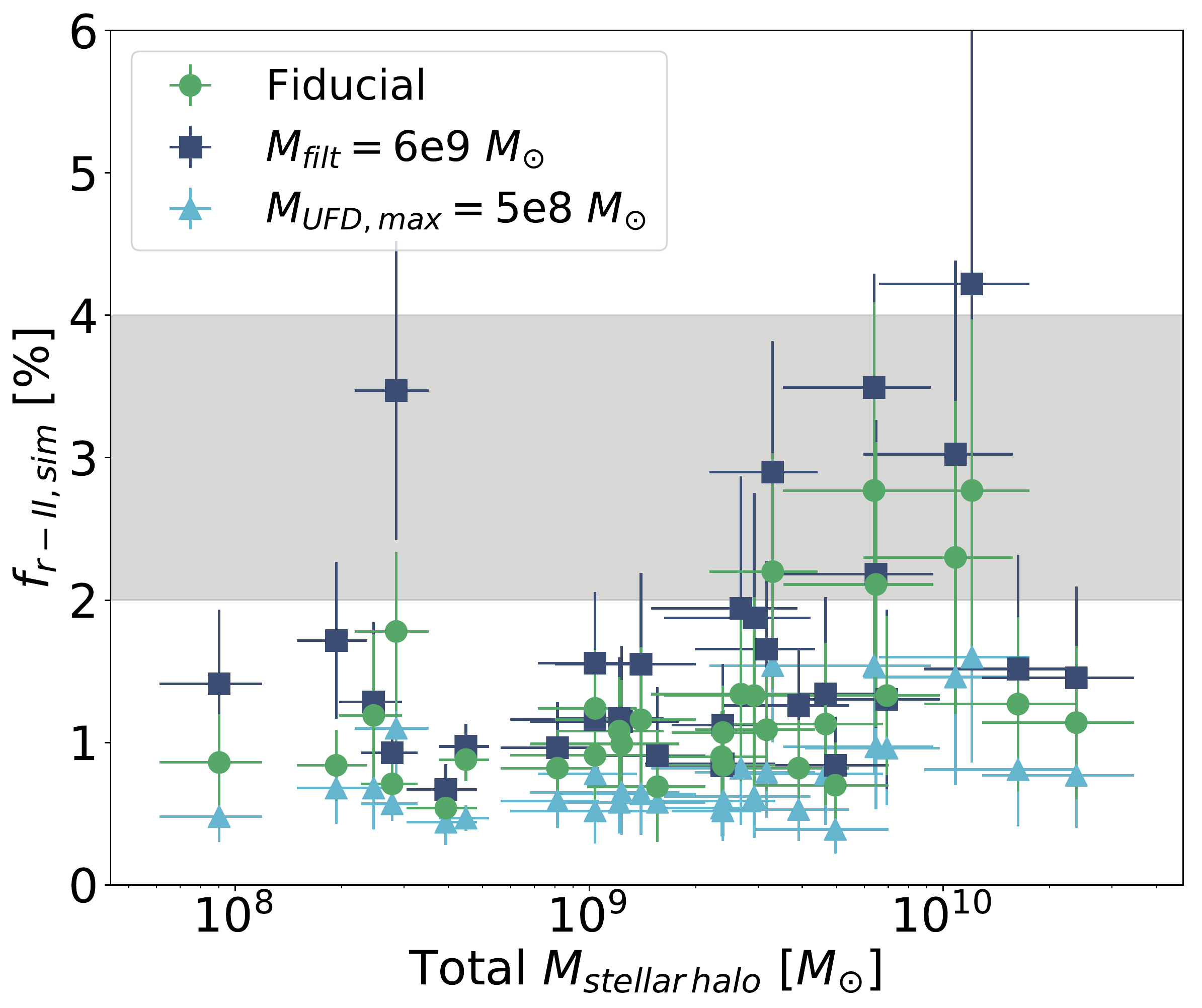}
\caption{The simulated r-II fraction for the individual \textit{Caterpillar} halos as a function of their total accreted stellar mass. Three models are shown as examples: the fiducial model, a variation of $M_{filt}$, and a variation of $M_{UFD,max}$. As in Figure \ref{fig:fracs}, the currently observed fraction of r-II stars in the Milky Way stellar halo is shown in grey. The error bars correspond to uncertainty in the empirical relations. There is a tendency for stellar halos with more accreted stellar mass to have a higher r-II fraction.
\label{fig:frac_Mstar}}
\end{figure}

We note that there appears to be a tendency for more massive stellar halos (stellar halos which formed from more massive destroyed galaxies) to have a higher $f_{r-II,sim}$. Figure \ref{fig:frac_Mstar} shows this. Because the Milky Way stellar halo is on the lower mass end of the range of \textit{Caterpillar} stellar halos (see Figure \ref{fig:ZvsM}), the simulated r-II fraction is slightly lowered if we only consider the six stellar halos with masses closest to the MW halo: $f_{r-II,sim,MW} \sim 1-1.5$\% for the fiducial model. This apparent trend could also be due to the large scatter, though.


\subsection{Fraction of Stars from Now-Destroyed UFDs}

\begin{figure*}
\gridline{\fig{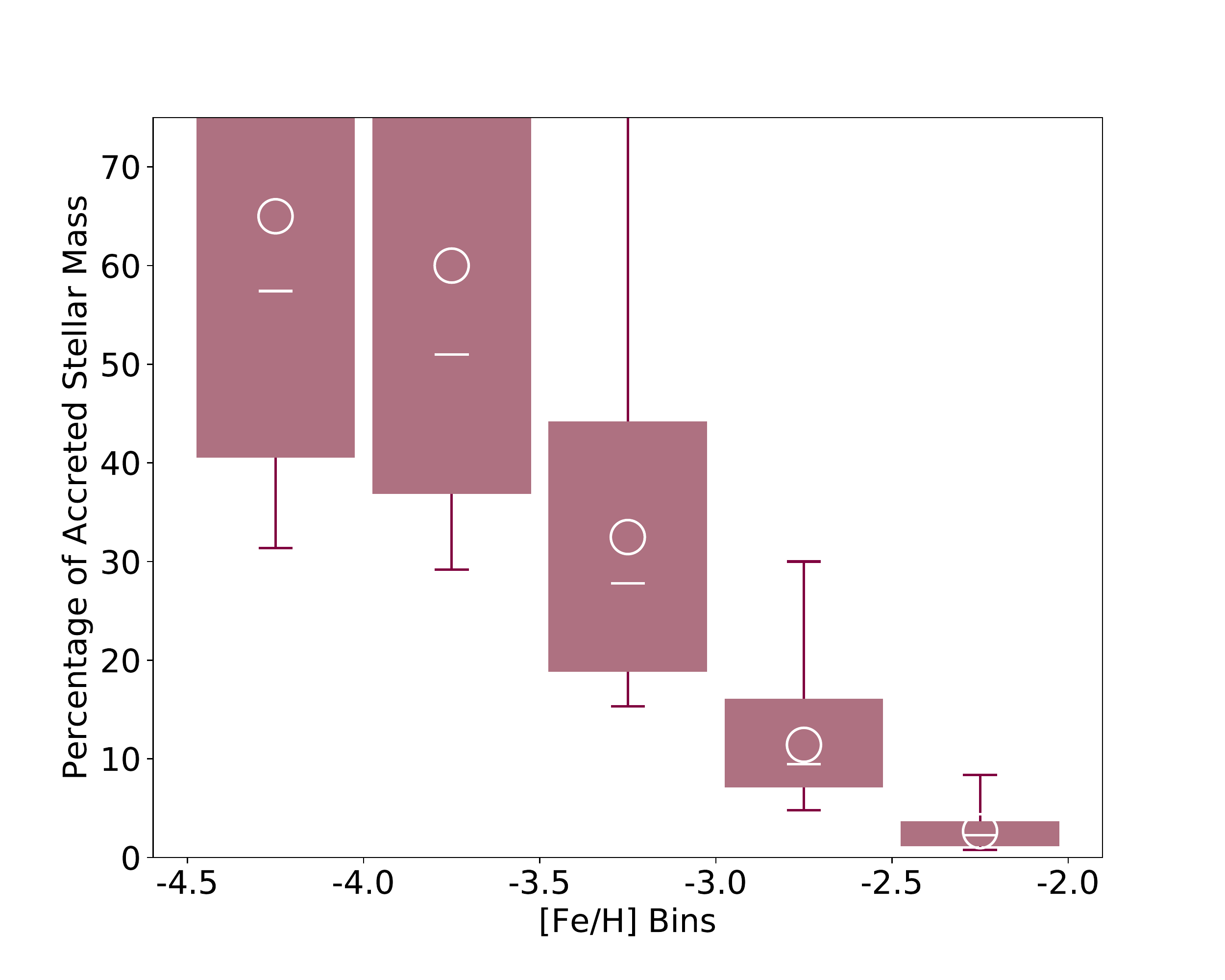}{0.5\textwidth}{(a) Accreted stellar mass that originated in UFDs}
	\fig{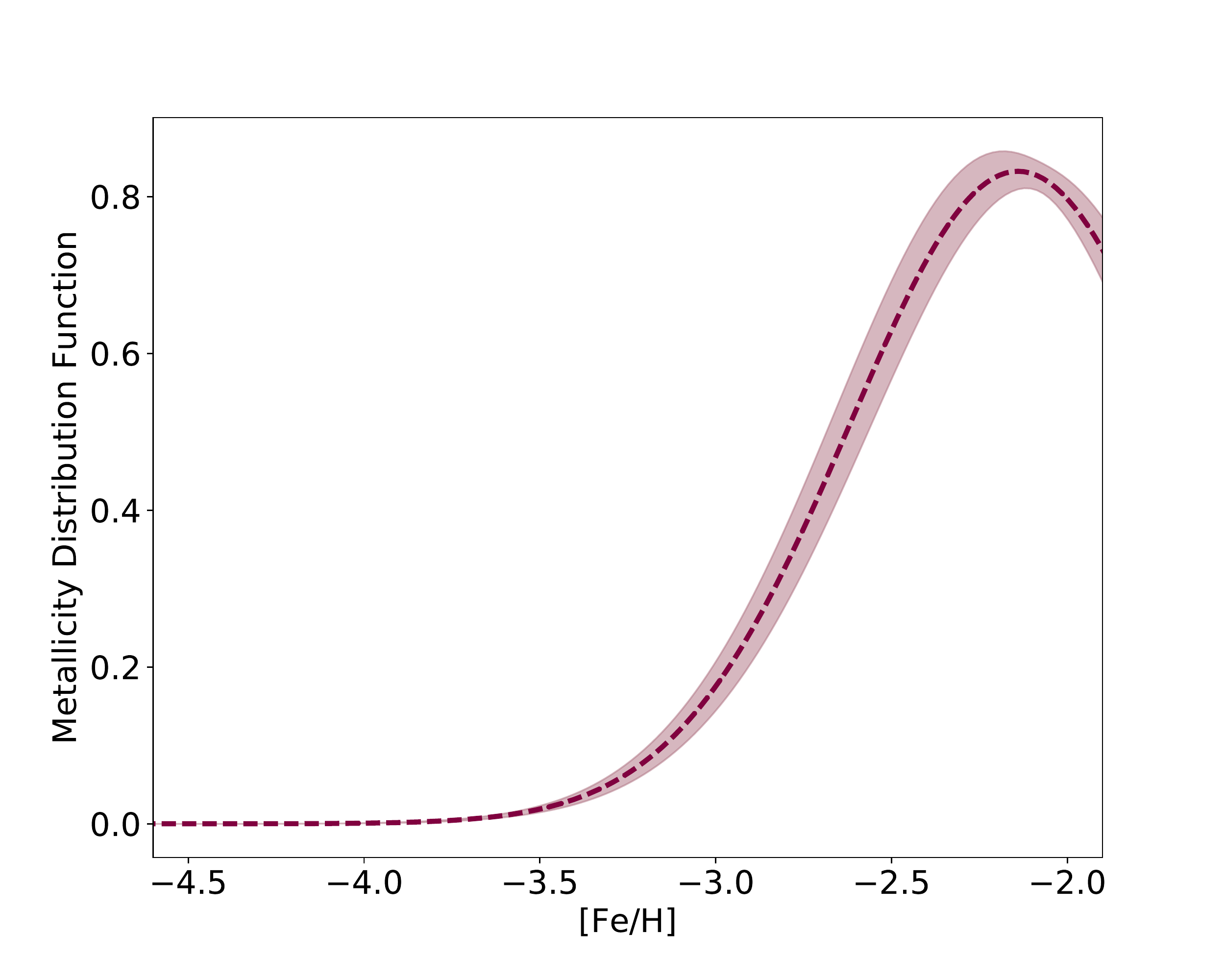}{0.5\textwidth}{(b) MDF of destroyed UFDs}}
\caption{Left: The percentage of accreted stellar mass at different metallicities that originated in now-destroyed UFDs. The mean percentage is shown as a white circle and the median is shown as a white line. The colored boxes correspond to 68\% scatter between simulations, and the error bars shown the minimum and maximum percentages across simulations.
From $\mbox{[Fe/H]} = -2.5$ to $-2$, stars from UFDs make up a few percent of the stellar halo. Right: The averaged metallicity distribution function of now-destroyed UFDs. Shaded region shows 68\% scatter between simulations. Both plots show results from our fiducial model.
\label{fig:UFDfracs}}
\end{figure*}

In Figure \ref{fig:UFDfracs}, we plot the total fraction of accreted stellar mass at different metallicities that originated in now-destroyed UFDs, $r$-process enhanced or not. The figure also shows the metallicity distribution function of all of the now-destroyed UFDs averaged across all simulations.

If we assume an $r$-process event occurs in approximately 10\% of UFDs, approximately 90\% of the now-destroyed UFDs produced low neutron-capture stars (stars that exhibit low abundances of neutron-capture elements such as Sr and Ba). Low neutron-capture (low n-cap) could thus be another key signature to identify stars from now-destroyed UFDs. If low n-cap stars at intermediate and low metallicities come primarily from UFDs, the fraction of low n-cap stars in the halo should look roughly like the fractions shown in Figure \ref{fig:UFDfracs}a (multiplied by ${\sim}0.9$).

\begin{figure*}
\plotone{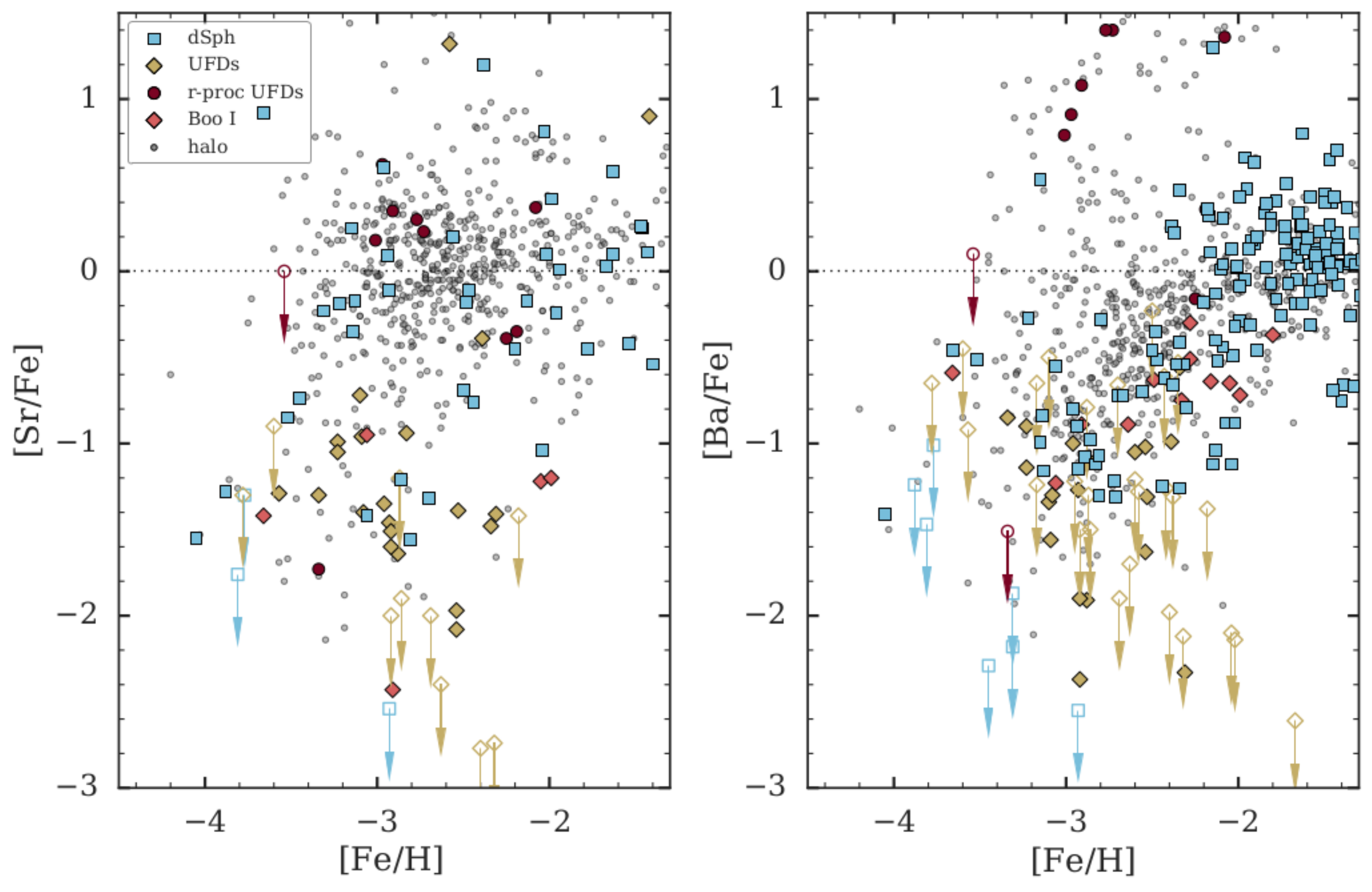}
\caption{Neutron-capture element abundances (Sr and Ba) for stars in surviving UFDs and in surviving dwarf spheroidal galaxies (dSph). The $r$-process UFDs (Reticulum II and Tucana III) and the UFD Bootes I are highlighted because they exhibit different behavior in their Sr and Ba abundances. Halo stars are shown in grey for comparison. The other 12 UFDs (shown in yellow) exhibit low [Sr/Fe] and [Ba/Fe] compared to the dSph (shown in blue). References are in Appendix \ref{sec:fig6cite}. \label{fig:lowncap}}
\end{figure*}

Figure \ref{fig:UFDfracs}b shows that the number of stars from now-destroyed UFDs peaks around $\mbox{[Fe/H]}\sim -2$. These stars make up only a few percent of the halo stars around this metallicity, though (Figure \ref{fig:UFDfracs}a). From $\mbox{[Fe/H]}\sim -2.5$ to $-2$, they are about as rare as r-II stars at low metallicities: $f_{low-ncap,sim} \sim 2-3$\%. While rare, finding low n-cap stars in this metallicity range could help us identify stars from now-destroyed UFDs. Based on observations of the Milky Way's satellite galaxies, below $\mbox{[Fe/H]}\lesssim -3$, low n-cap stars are found in both UFDs and more massive satellite galaxies, but from $\mbox{[Fe/H]}\sim -2.5$ to $-2$, UFDs appear to be the primary source.

Figure \ref{fig:lowncap} shows some of the neutron-capture element abundances ([Sr/Fe] and [Ba/Fe]) of stars in surviving UFDs relative to stars from the more luminous satellite galaxies around the Milky Way, the dwarf spheroidal galaxies (dSph). Halo stars are also shown in gray for comparison. Excluding the surviving UFDs that appear to have experienced an $r$-process event (Reticulum II and Tucana III), the UFD stars have lower Sr and Ba abundances than the dSph stars, most noticeably above $\mbox{[Fe/H]}\sim -3$. The UFD Bootes I is highlighted because it displays different behavior from other UFDs: its [Ba/Fe] ratios are higher and increase slightly with [Fe/H]. The other 12 UFDs clearly contain low n-cap stars relative to more luminous galaxies.

In the \citet{2014AJ....147..136R} sample of 313 metal-poor halo stars, the percentage of stars with $\mbox{[Ba/H]} < -3.5$ from $\mbox{[Fe/H]} = -2.5$ to $-2$ is 2.8\% and the percentage from $\mbox{[Fe/H]} = -3$ to $-2.5$ is 17.1\%, in rough agreement with Figure \ref{fig:UFDfracs}a.
As always, though, more observations are needed.
Upcoming halo star surveys without metallicity bias such as 4MOST and WEAVE \citep{4MOSTSPIE,WEAVESPIE} will expand on observations and allow us to study this question more in depth. For now, we merely note that low n-cap signatures are likely another key way to identify stars from now-destroyed UFDs, and we expect them to constitute a few percent of the stellar halo stars with $\mbox{[Fe/H]}\sim -2.5$ to $-2$.

\subsection{Number of Now-Destroyed $r$-Process Galaxies}

From the \textit{Caterpillar} simulations we can estimate the number of UFDs that merged to help form the Milky Way stellar halo. The \textit{Caterpillar} stellar halos are formed from $260 \pm 60$ UFDs on average. This includes both UFDs that merged directly into the host halo (roughly 1/3 of now-destroyed UFDs) and the UFDs that merged with other galaxies before merging with the host halo (roughly 2/3 of now-destroyed UFDs). Since ${\sim}10\%$ of UFDs appear to be $r$-process-enhanced, this means that ${\sim}20-30$ $r$-process UFDs may have contributed directly (${\sim}10$) or indirectly (${\sim}20$) to our stellar halo.

\subsection{Limitations} \label{subsec:lim}

There are potential issues with directly comparing \textit{Caterpillar} stellar halos to the Milky Way stellar halo. As mentioned in Section \ref{subsec:compare}, we are conflating ``accreted stars" with ``stellar halo", but actual stellar halos are not exclusively and comprehensively composed of accreted material. In our analysis, we do not consider \emph{in situ} stars. For simplicity's sake, we also do not consider that accreted stars can end up in the disk/bulge \citep[as in, e.g.,][]{Gomez17}. If a large portion of the metal-poor stellar halo originated \textit{in situ} or a large portion of the accreted material ended up in the disk, it would significantly affect the r-II fraction. The fraction of halo stars that formed \textit{in situ} can be large, and it is unclear how large \citep{Monachesi18}. \textit{In situ} halo stars are more metal-rich than accreted stars \citep{Bonaca17}, however, so metal-poor halo stars --- the focus of our analysis --- appear to be largely accreted. This is supported by both observations and hydrodynamics simulations \citep{Cooper15,Bonaca17,ElBadry18}. \citet{Bonaca17} kinematically identified accreted and \textit{in situ} halo stars in the Gaia DR1 + RAVE-on catalog, finding a bimodality about $\mbox{[Fe/H]}=-1$ with the accreted stars being more metal-poor than the \textit{in situ} stars. Their interpretation is supported by the Latte simulation from the FIRE project. Using more FIRE simulations, \citet{ElBadry18} found that ${\gtrsim}80$\% of the stellar halo stars below $\mbox{[Fe/H]} \sim -2.5$ are accreted (see their Figure 7). If none of the \textit{in situ} stars are highly $r$-process enhanced, this would increase $f_{r-II,obs}$ relative to $f_{r-II,sim}$ by at most a factor of ${\sim}1.2$. Given the uncertainty around the fractions and the likely overestimation of $f_{r-II,obs}$ due to observational bias, though, this would not change our finding that around half of r-II halo stars formed in now-destroyed UFDs.


We also do not consider the r-II stars that originate \textit{in situ} \citep[as in, e.g.,][]{Shen15,vanDeVoort15,Naiman18} or in more massive dwarfs. This is because we are specifically interested in how many of the observed r-II stellar halo stars may have originated in the low-mass UFDs. In theory, adding up the r-II fractions that result from each of these different possible r-II star channels should add up to 100\% of the observed fraction. Simulating these more complex r-II origins requires more sophisticated modeling than what is in the scope of this paper, however, so this will have to be investigated more in future work. The observed fraction must also be determined more accurately to better determine the overlap of $f_{r-II,sim}$ and $f_{r-II,obs}$. Furthermore, depending on the definition of ``ultra-faint dwarf," the r-II fraction from now-destroyed UFDs will be different. By considering only UFDs that can become strongly enriched from a single NSM, we limit ourselves to only the contributions from low-mass UFDs. This is a conservative choice, and if we included the contributions from r-II origin channels in more massive UFDs (e.g., r-II stars can form in more massive halos that experience more than one NSM event or inhomogenous mixing), the r-II fraction from UFDs would increase.

Additionally, we assume that $f_{NSM} \sim 10$\% of UFDs experience a NSM event (or other prolific $r$-process event), but this fraction is based on a small number of known UFDs. It may also make sense to determine $f_{NSM}$ in terms of total stellar mass that has been enriched by a prolific $r$-process event. Reticulum~II and Tucana~III are on the lower stellar mass end of UFDs, so weighting by stellar mass when determining $f_{NSM}$ significantly lowers the contribution of now-destroyed UFDs to r-II halo stars. In this case, $f_{NSM} \sim 2-3$\% and the simulated r-II fraction ($f_{r-II,sim} \sim 0.3$\%) would only account for around ${\sim}10$\% of the observed fraction.

Whether we determine $f_{NSM}$ by the number of UFDs that experience an $r$-process event or by the amount of stellar mass enriched by an $r$-process event is dependent on whether the NSM rate is dominated by a retention fraction or a production rate. If the NSM rate is dominated by a retention fraction, it would depend more on the total halo mass than the stellar mass. Because UFDs are in roughly the same halo mass range \citep{Strigari08,Jethwa18}, determining $f_{NSM}$ as we did in Section \ref{subsec:fNSM} should be more appropriate than weighting by stellar mass. On the other hand, if the NSM rate is dominated by a production rate, weighting by stellar mass is likely more appropriate. We note that the current LIGO rate is ten times higher than what is needed to produce all the $r$-process material in the Milky Way \citep{Ji2018,Belczynski18,LIGOGW170817a,2018ApJ...855...99C}, suggesting that the retention fraction is likely dominant and our determination in Section \ref{subsec:fNSM} is more appropriate. This remains uncertain for now, however. Future LIGO measurements will give clarity to this.

Furthermore, the [Fe/H] distributions have a fixed, simplistic shape. The individual Gaussians representing each destroyed halo have a physically motivated standard deviation and are able to reproduce similar bulk properties to those of observed stellar halos \citep{Deason16}, but we know they are not the true distributions. For example, the cumulative distribution functions of the  \emph{Caterpillar} stellar halos differ from that of the Milky Way stellar halo, particularly at the very lowest metallicities. The Gaussian [Fe/H] distributions used in this analysis are thus a simple choice to produce reasonable stellar halo MDFs, but they are insufficient to completely capture the true distribution of the Milky Way stellar halo and its satellites. We use the Gaussians in this analysis because we are unable to find a physically motivated distribution that better matches observations.

Additionally, our choice of $M_{star}-M_{peak}$ relation affects our results for $f_{r-II,sim}$. Using the \citet{GK17} relation (GK17) or \citet{GK14} relation (GK14) produces nearly the same r-II fractions, but using the \citet{Moster13} relation produces r-II fractions that are roughly half as large. Using the \citet{Behroozi13} relation more than doubles the r-II fraction (producing unreasonably high fractions), and using the \citet{Brook14} relation gives fractions that are roughly one-fourth of those produced by GK17 or GK14. The disagreement between these different relations displays the uncertainty that abundance matching relations have regarding low-mass halos such as UFDs. We focus on the most up-to-date $M_{star}-M_{peak}$ relation, GK17, but the potential issues with using abundance matching relations to assign mass to low-mass halos should be kept in mind.

If future work continues to use empirical relations, the work could potentially be improved by using a [Fe/H] distribution with a more pronounced metal-poor tail. Having used empirical relations here to obtain an initial idea of what is reasonable in our model, however, we believe semi-analytic modeling will provide a better avenue for future investigation into this and similar questions.

Lastly, subhalos passing close to the host galaxy's center should probably be destroyed by the host galaxy's disk, but are not because the \textit{Caterpillar} simulations are dark matter only \citep{GK17b}. Including surviving subhalos in the stellar halo does not significantly change the r-II fractions, though, so this does not appear to be significant to our results on r-II fractions.

\section{Conclusions} \label{sec:conc}

We investigate the possibility that highly $r$-process enhanced metal-poor stars (metal-poor r-II stars) largely originated in the smallest, earliest galaxies (early analogs of ultra-faint dwarfs, UFDs) that merged into the Milky Way over the course of its formation history. Our results support this possible connection between r-II stars and the smallest building blocks of our galaxy. 
We find that around half of r-II stars
may have originated in now-destroyed ultra-faint dwarfs that experienced a rare prolific $r$-process event such as a neutron star merger.

We reach this conclusion by simulating what fraction of low-metallicity stellar halo stars could have become highly $r$-process enhanced in now-destroyed UFDs. This fraction is the simulated r-II fraction, $f_{r-II,sim}$. We compare this to the observed r-II fraction, the fraction of low-metallicity stellar halo stars that have been observed to be highly $r$-process enhanced. Assuming the most likely values for parameters in our model ($z_{reion}\sim 8$, intermediate mass thresholds, $f_{NSM}\sim 10$\%) gives a simulated $f_{r-II,sim} \sim 1-2 \%$, accounting for around half of the observed $f_{r-II,obs} \sim 2-4 \%$. In cases where we choose the most extreme parameter values, $f_{r-II,sim}$ ranges from ${\sim}0.01-4$\%. Considering scatter between simulations and less extreme variation of model parameters, $f_{r-II,sim}$ can account for ${\sim}20-80$\% of $f_{r-II,obs}$. Due to incomplete sampling, though, $f_{r-II,obs}$ likely overrepresents the fraction of r-II halo stars. This means the percentage of $f_{r-II,obs}$ that $f_{r-II,sim}$ can account for is likely closer to ${\sim}80$\% than ${\sim}20$\%.

To determine the simulated $f_{r-II}$, we use high-resolution dark-matter cosmological simulations (the \textit{Caterpillar} suite), empirical relations linking dark matter mass to stellar mass and metallicity, and a simple, empirically motivated $r$-process treatment. Our $r$-process treatment assumes that $5-15$\%, or ${\sim}10$\%, of early UFDs experience an early prolific $r$-process event that enriches all of the gas from which their subsequent stars form with $r$-process elements. The $r$-process event is most likely a neutron star merger, but the model is agnostic to the specifics of the event. The ${\sim}10$\% fraction comes from the fraction of surviving UFDs that have been observed to be $r$-process enhanced.

Intriguingly, there is some recent evidence that $r$-process-enhanced stars may have kinematics associated with accretion.
Abundances of high-velocity stars in Gaia DR1 \citep{HerzogArbeitman18} and Gaia DR2 \citep{Hawkins18} have found 2/10 such stars appear to have $\mbox{[Eu/Fe]} > 1$, a much higher fraction than is found for random metal-poor stars in the halo. The high velocities suggest these stars originate in accreted satellites.
Additionally, \citet{Roederer18} recently studied the kinematics of all known $r$-process-enhanced stars in Gaia DR2, also finding evidence that these stars appear to have an accretion origin from UFDs or low-luminosity classical dwarf spheroidals.
The statistics in these studies are still low, 
but they support our hypothesis of an accretion origin for $r$-process-enhanced stars. The kinematics of r-II Milky Way halo stars are currently being studied in more detail (e.g., Ji et al. in prep).  


Stars with low abundances (or no detection) of $r$-process elements (low neutron-capture stars, or low n-cap stars) could be another way to identify stars that originated in now-destroyed UFDs. If an $r$-process event occurs in ${\sim}10$\% of UFDs, ${\sim}90$\% of UFDs should produce low n-cap stars. Our model predicts that ${\sim}2$\% of the halo stars with $\mbox{[Fe/H]} = -2.5$ to $-2$ should be low n-cap stars from UFDs. This is in rough agreement with the sample of metal-poor halo stars from \citet{2014AJ....147..136R}, but more data from upcoming halo star surveys such as 4MOST and WEAVE will allow this to be studied more in depth.

There are a number of limitations in this model, including how we determine $f_{NSM}$ and the imperfections of the empirical relations. Future work on predicting the actual number of r-II halo stars or their distribution in the stellar halo will require more detail than we go into here. The results of this initial investigation, however, support a strong connection between metal-poor r-II stars and now-destroyed UFDs. 
Neutron-capture element abundances of Milky Way halo stars may thus allow us to quantify how much these small, relic galaxies contribute to the formation of our Galaxy.

\acknowledgments

We greatly thank Brendan Griffen for running the \textit{Caterpillar} simulations and providing helpful comments.
We thank Antonela Monachesi for providing data on galaxies in the GHOSTS survey, and we thank Ian Roederer for useful discussions.
K.V.B. was supported by the Whiteman Fellowship at MIT and the United States Department of Energy grant DE-SC0019323 while conducting this research.
A.P.J. is supported by NASA through Hubble Fellowship grant HST-HF2-51393.001 awarded by the Space Telescope Science Institute, which is operated by the Association of Universities for Research in Astronomy, Inc., for NASA, under contract NAS5-26555. A.F. acknowledges support from NSF grants AST-1255160 and AST-1716251.
F.A.G. acknowledges financial support from FONDECYT Regular 1181264, and funding from the Max Planck Society through a Partner Group grant.
B.W.O. was supported by the National Aeronautics and Space Administration (NASA) through grants NNX15AP39G, 80NSSC18K1105, and Hubble Theory Grant HSTAR-13261.01-A, and by the NSF through grant AST-1514700.  
This work benefited from support by the National Science Foundation under Grant No. PHY-1430152 (JINA Center for the Evolution of the Elements).

%

\vspace{5mm}


\software{\texttt{numpy} \citep{numpy}, \texttt{scipy} \citep{scipy}, \texttt{matplotlib} \citep{matplotlib}}



\appendix

\section{Figure 6 References} \label{sec:fig6cite}
\citet{AOK07b,AOK09,COH09,COH10,Francois16,2010AN....331..474F,Frebel10b,Frebel14,Frebel16,FUL04,GEI05,Hansen17,Hansen18,JAB15,KIR12,Kirby17,SHE01,SHE03,Simon10,SIM15,SKU15,TAF10,TSU15,TSU17,URA15,Chiti18,Feltzing09,GIL13,LAI11,ISH14,Ji16b,Ji16a,Ji18prep,Roederer14a,Roederer16b,Koch08,Koch13,Koch14b,Nagasawa18,Norris10b,NOR17,VEN12,Venn17}




\bibliography{simbib,alexbib,kaleybib}





\end{document}